\documentclass[cits]{PoS}
\usepackage{amsmath}
\usepackage{multirow}
\usepackage{wrapfig}
\title{The observation possibility of $B_c$ excitations at LHC }

\ShortTitle{The observation possibility of $B_c$ excitations at LHC }

\author{\speaker{Aleksander Berezhnoy}\\%
        SINP MSU, Moscow, Russia\\
        E-mail: \email{Alexander.Berezhnoy@cern.ch}}

\author{Anatolii Likhoded\\
       IHEP, Protvino, Russia\\
       E-mail: \email{Anatolii.Likhoded@ihep.ru}}

\abstract{ 
 It is shown that in the decays $B_c(2P)\to B_c^* \gamma^\mathrm{hard}$, $B_c(3P)\to B_c^* \gamma^\mathrm{hard}$ and $B_c(2S)\to  B_c^* +\pi^+\pi^-$ followed by the decay $B_c^*\to B_c +\gamma^\mathrm{soft}$ the loss of the soft photon $\gamma^\mathrm{soft}$   do not wash out peaks from the  initial excitations. The relative yields of $B_c^*$, $2P$-wave, and $3P$-wave  states of $B_c$ meson at LHC are estimated as function of transverse energy of emitted photon $\gamma^\mathrm{hard}$. It is pointed out, that the decays $B_c(2S) \to B_c(B_c^*)+\pi^+\pi^-$ could provide a new information about a $\sigma$ resonance nature. }

\FullConference{The XXI International Workshop  High Energy Physics and Quantum Field Theory \\
		 June 23 -- June 30,  2013 \\
		 Saint Petersburg Area, Russia}

\begin{document}

\section{Introduction}
The first observation of $B_c$-meson ground state was performed by CDF and D0 experiments (Tevatron) 
in two decay modes: $B_c \to J\psi l \nu$ ($l = e, \mu$) and $B_c \to J\psi \pi$~\cite{Aaltonen:2007gv,Abazov:2008rba,Abazov:2008kv,Abe:1998wi}.  For now, the $B_c$ meson  is observed also by LHC experiments in the following decay modes: $B_c \to J\psi \pi$ (LHCb, CMS, ATLAS~\cite{Aaij:2012dd, CMS:2012oxa,ATLAS:2012bja}),  $B_c \to J\psi \pi \pi \pi$ (LHCb and CMS~\cite{CMS:2012oxa,LHCb:2012ag}), $B_c^+ \to \psi(2S)\pi^+$,  $B^+_c \rightarrow J/\psi D_s^+$ and $B^+_c \rightarrow J/\psi D_s^{*+}$, $B_{c}^{+}\to J/\psi K^+$ (LHCb~\cite{Aaij:2013oya,Aaij:2013gia,Aaij:2013vcx}).

$B_c$-meson is a nonrelativistic doubly heavy system, and like $b \bar b$- and $c \bar c $-states, 
can be considered within the potential model, which predicts 19 bounded $b\bar c$-states 
below the decay threshold to $B\bar{D}$ (see \cite{Gershtein:1994jw,Gershtein:1997qy, Gouz:2002kk,Godfrey:2004ya}). 
Contrary to  $b\bar b$- or $c \bar c$-quarkoniums, there are no strong  annihilation decay modes for  $b \bar c$-states, and this makes the $b\bar c$ system similar to usual $b\bar q$ or $c\bar q$-meson.
All excited $B_c$-mesons decay via cascade electromagnetic or hadronic transitions to the ground state. 

There are some differences in the production of ordinary quarkonium with hidden flavor and the production of $B_c$-meson. The necessity to produce two heavy quark pairs, $b\bar b$ and $c\bar c$,  leads to the strong suppression of  $B_c$ production cross section:

\begin{equation}
\sigma_{B_c} \sim 10^{-3} \sigma_{B}.
\end{equation}

It is natural, that  the difference in  production mechanisms 
leads to the difference in the relative yield of excited  states.
Contrary to the production of quakonia with hidden flavor, the production of $P$-wave $B_c$ states is suppressed comparing to $S$-wave states. The mechanism of $B_c$ states  production was studied theoretically in~\cite{Berezhnoy:1994ba,Chang:1994aw, Berezhnoy:1995au,Kolodziej:1995nv,Berezhnoy:1996ks,Berezhnoy:1997fp,Baranov:1997sg, Baranov:1997wv, Berezhnoy:1997uz, Chang:2004bh, Berezhnoy:2004gc,Chang:2005wd,Chang:2006xka,Berezhnoy:2010wa}. 
According to these investigations the $B_c$-meson production can be interpreted as a $b$-quark fragmentation (like fragmentation $b \to B$) only 
at high transverse momenta ($p_T > 35~\mbox{GeV}$). The ratio $R_{B_c} = \sigma(B_c^*)/\sigma(B_c)$ between production cross sections of vector and pseudoscalar ground states predicted within the fragmentation approach is about $ 1.4$.
At low and middle $p_T$ the recombination mechanism 
dominates and gives $R_{B_c} \sim 2.6$.  Thus, the value $R_{B_c}$ could provide the essential information about the production mechanism.
  Unfortunately, as it shown in this paper, the experimental measurement  of $R_{B_c}$   is currently impossible due to the low energy release in $B_c^* \to B_c \gamma$ decay, which leads to the essential difficulties  in the $B_c^*$ identification.

The analysis below shows, that electromagnetic  transitions $B_c(2P) \to B_c(1S) \gamma$ and hadronic transitions $B_c(2S) \to B_c(1S) \pi \pi$ are the more perspective for the experimental observation, than  the process $B_c^*\to  B_c \gamma$, in spite of that the total yield of $B_c^*$ is several times lager.

\section{Electromagnetic decays of $B_c^*$ and $P$-wave $B_c$ states}

The mass difference between the lowest vector and pseudoscalar states of $\bar b c$-quarkonium ($B_c^*$ and $B_c$) predicted within potential models  is fairly small:
\begin{equation}
M(B_c^*)- M(B_c)\approx 65~\mbox{MeV}. 
\end{equation}
 It is practically impossible to detect the photon with the transverse energy  about $65$~MeV at LHC experiments.  Even the LHCb experiment, which is designed for low $p_T$ physics, can not detect a photon with the transverse energy smaller, than $300$ MeV approximately. Therefore the decaying $B_c^*$ must have a quite large transverse momentum to be observed. 
The production cross section rapidly decreases  with increasing of the transverse momentum, and this leads to significant decreasing of  event number,  where such a photon could be detected.

According to the expression for the maximum transverse energy of photon
\begin{equation}
\omega_T^{max} = \left( 1+\frac{\Delta M}{2M_{B_c^*}}\right) \Bigl( \sqrt{M_{B_c^*}^2+p_T^2}+p_T\Bigr)\frac{\Delta M}{M_{B_c^*}} 
\approx 0.01\Bigl( \sqrt{M_{B_c^*}^2+p_T^2}+p_T\Bigr) 
\label{eq:boosted_photon}
\end{equation}
the photons with transverse energy $\omega_T> 0.5$~GeV can be produced  by $B_c^*$-mesons with $p_T(B_c^*)> 24$~GeV. 
Such a cut on transverse momentum reduces the yield of $B_c^*$-mesons approximately by two orders of magnitude. It should be noted that 
there is no such a problem for $2P$-wave $B_c$ states, at least for LHCb experiment. 
Even  $B_c(2P)$-mesons with low transverse momentum are able to produce the fairly hard photon (see tab.~\ref{tab:photon_decays}). Therefore, despite  the low relative yield 
of $2P$-wave excitations (about 10-20\%  v.s. $\sim$60\% for $B_c^*$), the $2P$-wave excitations need smaller efforts to be detected. 
Rough estimations show, that a number of photons with $\omega_T> 0.5$~GeV from $P$-wave states is $25\div 50$ times larger, than from  $B_c^*$ mesons: 

\begin{equation}
\frac{\sigma_{2P}(\omega^{\gamma}_{T} > 0.5~\mathrm{GeV})}{\sigma_{1S} (\omega^{\gamma}_{T} > 0.5~\mathrm{GeV}   \Longleftrightarrow  p_T(B_c)>24~\mathrm{GeV})}\sim 25\div 50.
\end{equation}

It should be stressed that only about 20\% of all $2P$-wave states emit only one photon, immediately transforming to the lowest pseudoscalar state:
$$B_c(2P 1^+)\xrightarrow[\sim 13 \%]{\gamma} B_c(1 ^1S_0),$$
$$B_c(2P 1^{'+})\xrightarrow[\sim 94 \%]{\gamma} B_c(1 ^1S_0). $$

In all other cases $2P$-wave excitations  decay through  intermediate $B_c^*$ state,  consequentially emitting a ``hard''  photon and a ``soft'' one:
$$B_c(2P) \xrightarrow{\gamma^{\textrm{hard}}} {B_c(1 ^3S_1)}  \xrightarrow{\gamma^{\textrm{soft}}} {B_c(1 ^1S_0)}.$$

The second (``soft'') photon will be lost almost always. However, this  doesn't prevent the experimental observation of $2P$-wave states of $B_c$-meson. 
Indeed, it can be easily shown, that  the loss of ``soft'' photon in the cascade decay of $2P$-wave states broadens  the peak by the value 
\begin{equation}
\Delta\tilde{M} \approx 2 \frac{\Delta M \Delta M'}{M}
\label{eq:width}
\end{equation}
 and shifts it to the left by $\Delta M$, where  $\Delta M= M(B_c^*)-M(B_c)$  and $\Delta M'= M(B_c^P)-M(B_c^*)$.
 
The numerical value of  broadening for $2P$-wave states can be obtained from~(\ref{eq:width}) by putting  $\Delta M\approx 65$~MeV and $\Delta M'\approx 400$~MeV:
$$\Delta\tilde{M}_{2P} \approx 10~\textrm{MeV}.$$

\begin{table}[!t]
\centering
\parbox[t]{0.47\textwidth}{
\caption{ Radiative decays  of $B_c$ meson $P$-wave excitations (see~\cite{Godfrey:2004ya,Gupta:1995ps,Kiselev:1994rc}). \label{tab:photon_decays} \hfill}}
\hfill
\parbox[t]{0.47\textwidth}{
\caption{ The relative yield of  excited $B_c$ states in the decay mode $B_c+\gamma$  as function of a minimal transverse energy of photon $\omega_T^{\textrm{min}}$. \label{tab:Bc_photon_yield} \hfill}}
\\
\resizebox*{1.0\textwidth}{!}{
\begin{tabular}{|c|c|c|c|}
\hline
initial state & final state & Br, \% & $\Delta M$, MeV \\
\hline 
$2^3P_0$ & $1^3S_1+\gamma$ &  100 &  363-366 \\
$2 P 1^+$ & $1^3S_1+\gamma$ &  87 &  393-400 \\
 & $1^1S_0+\gamma$ &  13 &  393-400 \\
$2P 1'^+$ & $1^1S_0+\gamma$ &  94 &  472-476 \\
  & $1^3S_1+\gamma$ &  6  &  472-476 \\
$2^3P_2$ & $1^3S_1+\gamma$ &  100 &  410-426 \\
\hline
$3^3P_0$ & $1^3S_1+\gamma$ & 2  &  741 \\
$3 P 1^+$ & $1^3S_1+\gamma$ &  8.5 &  761 \\
 & $1^1S_0+\gamma$ &  3.3 &   820 \\
$3P 1'^+$ & $1^1S_0+\gamma$ &  22.6 &  825 \\
  & $1^3S_1+\gamma$ &  0.7 &  769 \\
$3^3P_2$ & $1^3S_1+\gamma$ &  18 &  778 \\
\hline
\end{tabular}
\hfill
\begin{tabular}{|c|c|c|}
\hline
$\omega_T^{\textrm{min}}$, GeV & $B_c$ state & relative yield \% \\
\hline
\multirow{3}{*}{0.3} &   $B_c(2P)$ & $\sim 5.0$ \\
 &  $B_c(3P)$ & $\sim 1.0$ \\
 & $B_c^*$ & $\sim 0.8$ \\ \hline
  \multirow{3}{*}{0.5} &   $B_c(2P)$ & $\sim 3.5$ \\
 &  $B_c(3P)$ & $\sim 0.7$ \\
 & $B_c^*$ & $\sim 0.06$ \\ \hline
\multirow{3}{*}{1.0} &   $B_c(2P)$ & $\sim 0.9$ \\
 &  $B_c(3P)$ & $\sim 0.4$ \\ 
 & $B_c^*$ & $\sim 0.005$ \\ \hline
\end{tabular}
}
\end{table}

It is clear, that this width is smaller than the apparatus width of resonance and doesn't affect the observation possibility at all. 
The broadening value for $3P$-wave states is also small:
$$\Delta \tilde{M}_{3P} \approx 20~\textrm{MeV}.$$

\begin{figure}[!t]
\centering
\resizebox*{1.0\textwidth}{!}{
\includegraphics{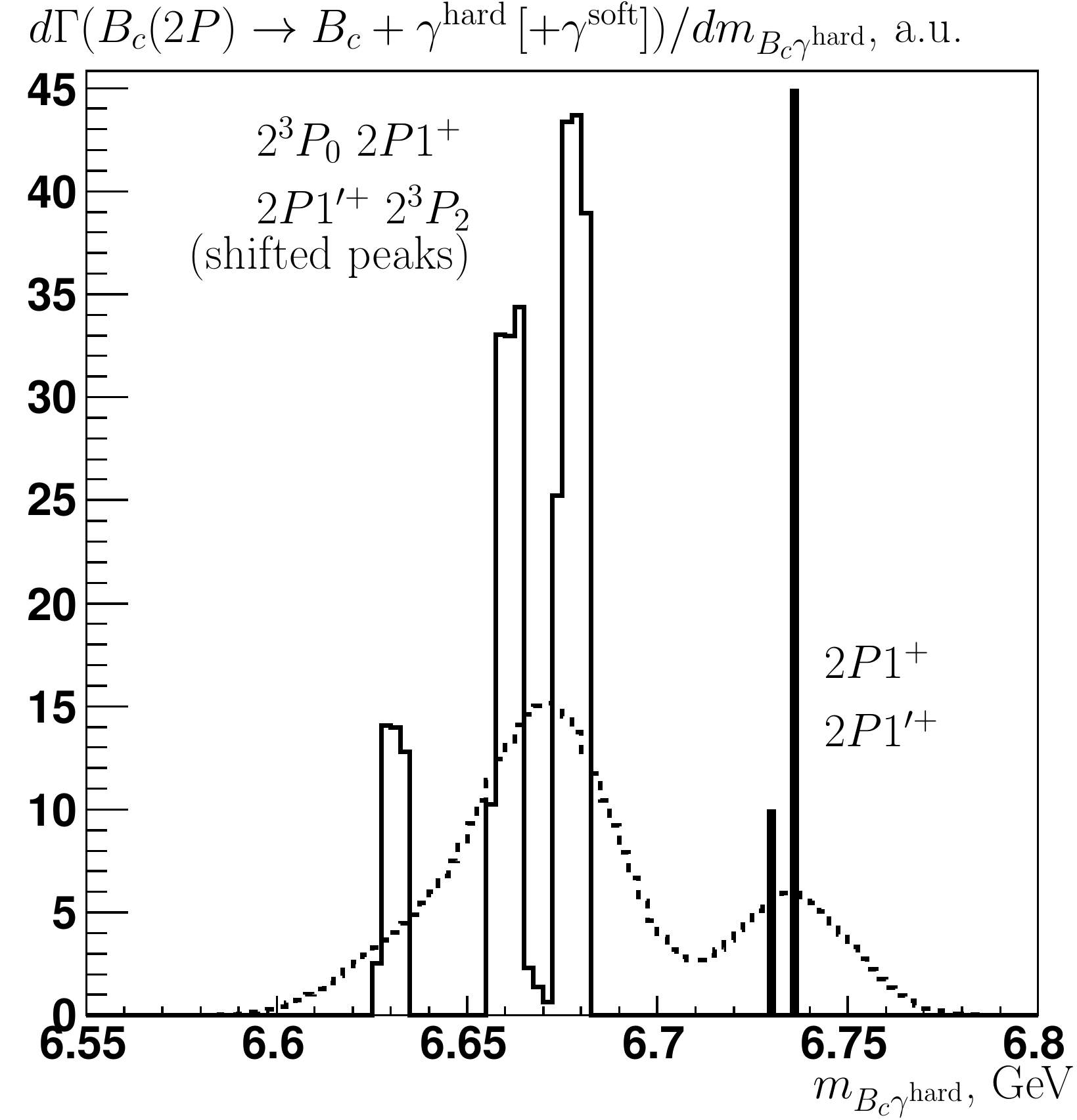}
\hfill
\includegraphics{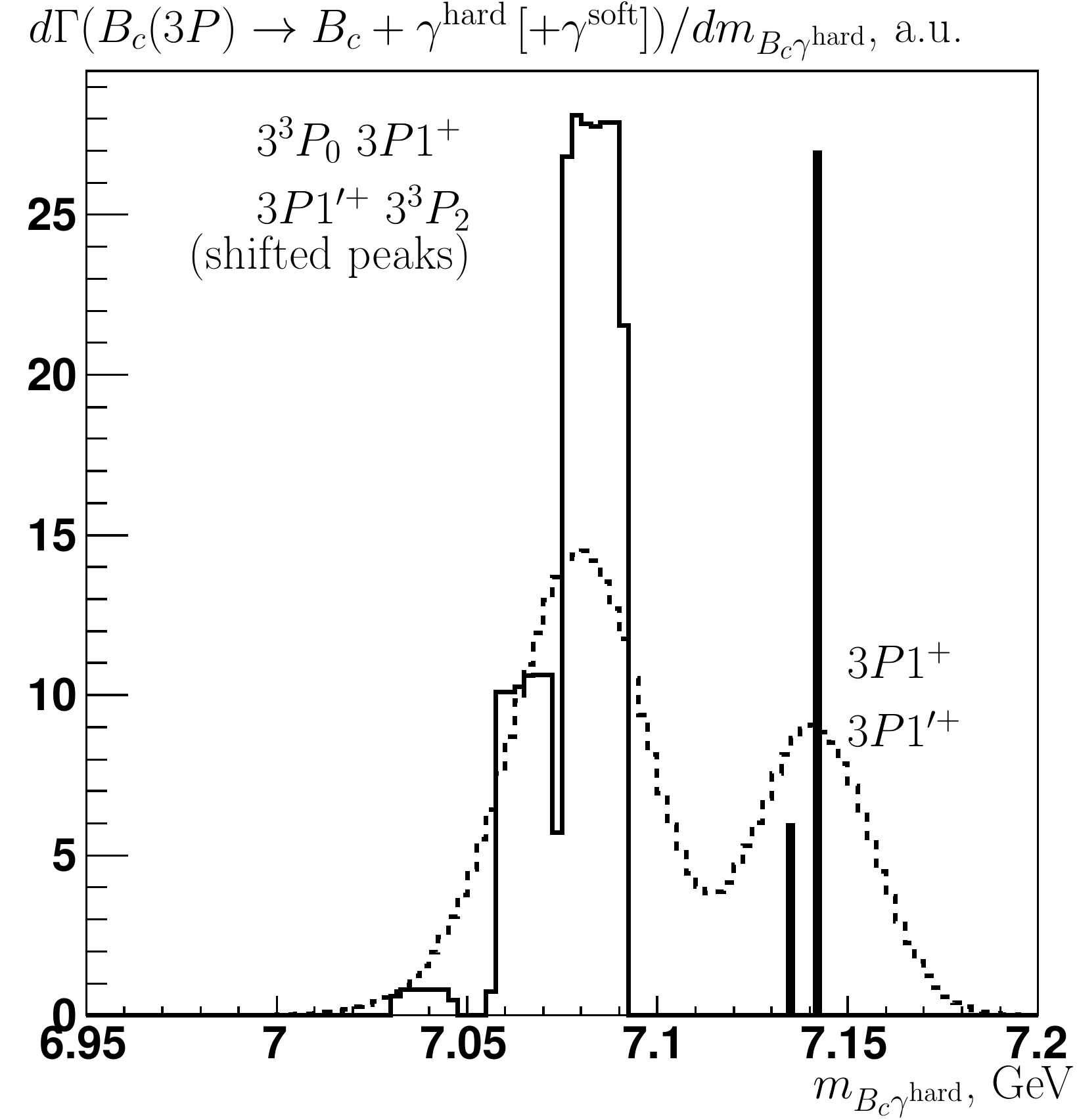}
}
\\
\parbox[t]{0.47\textwidth}{
\caption{  The mass spectrum of $B_c+\gamma^{\textrm{hard}}$ system  for the process $B_c(2P)\to B_c + \gamma^{\textrm{hard}} [+\gamma^{\textrm{soft}} ]$.}
\label{fig:2p}   \hfill}
\hfill
\parbox[t]{0.47\textwidth}{
\caption{The mass spectrum of  $B_c+\gamma^{\textrm{hard}}$ system  for the process $B_c(3P)\to B_c + \gamma^{\textrm{hard}} [+\gamma^{\textrm{soft}} ]$. \label{fig:3p}   \hfill}}
\end{figure}

The predicted distribution over $B_c+\gamma^{\textrm{hard}}$ invariant mass  for the cascade decays of $2P$-wave states is shown in fig.~\ref{fig:2p}. 
 Solid histograms correspond to the shifted and broadened 
peaks from $2^3P_0$, $2 P~1^+$, $2P~1'^+$ and $2^3P_2$. Unshifted peaks from $2P~1^+$ and $2P~1^{'+}$ states are conventionally marked by bold vertical lines.
To take into account the apparatus resolution, this distribution is convoluted with a Gaussian function with a dispersion $15$~GeV (see the  dashed histogram). Of cause,  such a simulation  doesn't reflect real properties of detector. Nevertheless, it gives a rough idea about shapes of peaks, which could be observed at LHC experiments.

Similar distributions for $3P$-wave states are shown in fig.~\ref{fig:3p}.

It is important to note that in spite of practically the same  yield of $2P$-wave and $3P$-wave states in the proton-proton interactions, the observation of $3P$-wave excitations in  $B_c+\gamma$ spectrum is more difficult. The thing is that $2P$-excitations always decays via electromagnetic transitions, while only  20\% of $3P$-wave states decay electromagnetically. 

The relative yields of $2P$-wave  and $3P$-wave states in  $B_c+\gamma$ decay mode for various minimal transverse energies of  photon are given in tab.~\ref{tab:Bc_photon_yield}.
It is seen from this  table, that the larger transverse energy of a photon, the smaller probability, that it comes from $B_c^*$-meson.

\section{$B_c(2S)\to B_c(1S)+\pi\pi$ decays}

There is every reason to expect, that in the long term the LHC experiments will allow us to observe not only $P$-wave states, but also $2S$-wave excitations of $B_c$-meson. 
Indeed, the yield of $2S$ is about 25\% of total $B_c$-meson yield and approximately a half of them decay to $B_c$ ($B_c^*$) and $\pi^+\pi^-$ pair:

$$B_c(2 ^1S_0)\xrightarrow[\sim 50 \%]{\pi^+\pi^-} B_c(1 ^1S_0),$$
$$B_c(2 ^3S_1)\xrightarrow[\sim 40 \%]{\pi^+\pi^-} B_c(1 ^3S_1),$$

$$\sigma(B_c(2S))/\sigma^{\rm total}(B_c) \sim 25~\%,$$
$$\sigma(2 ^3S_1)/\sigma(2 ^1S_0 ) \sim 2.6.$$

Therefore about 10\% of observed $B_c$-mesons originate from the decay $B_c(2S) \to B_c(B_c^*)+\pi^+\pi^-$.
 However, the separation of the signal from the background for such processes is a very complicated experimental task. In this paper we restrict ourselves to discussion about the signal shape, which could be observed after the cut-off of background. 
The only thing that we want to mention here, that the background conditions strongly depend on $B_c$ transverse momentum. Therefore it could be supposed, that the higher momentum will have a  $B_c(2S)$-meson, the more favorable  background conditions will be for that event. If it is so, the high $p_T$ experiments like CMS and ATLAS will be more suitable for the research of $B_c(2S)$, than LHCb. Of course the detailed estimations are needed to prove or disprove this hypothesis.

It would be very interesting to compare the experimentally measured spectra  of the $\pi\pi$-pair invariant mass  for decays   $B_c(2S) \to B_c + \pi^+\pi^-$ and $\psi' \to \psi + \pi^+\pi^-$.
The last one is investigated theoretically since 1970s. 
In~\cite{Brown:1975dz,Novikov:1980fa,Voloshin:1975yb,Voloshin:1980zf} it  was concluded  within the chiral theory, that at small $\pi\pi$  masses the process amplitude is approximately proportional to $m_{\pi\pi}^2-4m_{\pi}^2$. This approximation was extended to the total phase space of decay. Currently this simplest approach is used in LHCb for simulation of the discussed events. However, 
the data of BESII experiment~\cite{Ablikim:2006bz} shows, that  the contribution of  $\sigma$-resonance~\footnote{$\sigma$ or $f_0(500)$: $J^{PC}= 0^{+ +}$, $m=(400-550)-i(200-350)$ MeV.} to this process should also be considered. 
It is likely, that $B_c(2S)$-mesons emit the $\pi\pi$-pair via the same or similar mechanisms, as $\psi'$, therefore the role of $\sigma$-meson  could be also essential in the decays $B_c(2 ^3S_1) \to B_c^* + \pi^+\pi^-$ and $B_c(2 ^1S_0) \to B_c + \pi^+\pi^-$. 

It should be noted that there is no consensus about the nature of  $\sigma$-meson.  
The authors of~\cite{Albaladejo:2012te} consider the $\sigma$-meson as a dynamically generated resonance 
in $\pi\pi$ interactions. 
Within the approach developed in~\cite{Black:2000qq, Harada:2012km}  it is
a mixed state of two-quark and four-quark states. 

 The role of $\sigma$-meson in
$D_1(2430) \to D\pi\pi$ decay was studied in~\cite{Harada:2012km}.

\begin{wrapfigure}{r}{0.5\textwidth}
 \centering
\resizebox*{0.45\textwidth}{!}{\includegraphics{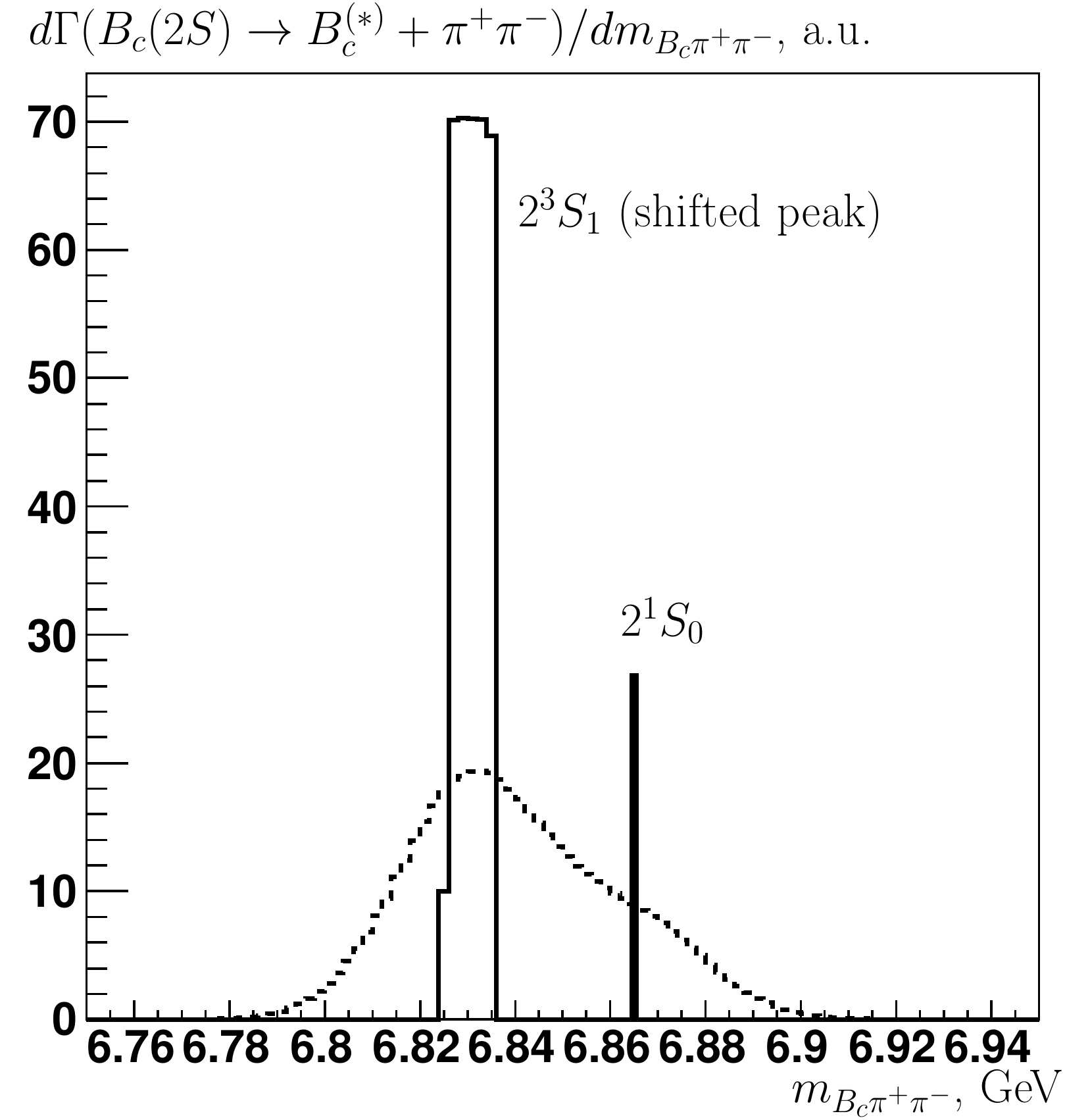}}
\caption{Mass spectrum of $B_c+\gamma^{\textrm{hard}}$ system for the process $B_c(2S)\to B_c + \pi \pi [+\gamma_{\textrm{lost}} ].$}
\label{fig:2s}
\end{wrapfigure}

As for the electromagnetic decays of $P$-wave excitations, the loss of ``soft'' photon in $B_c^*$  decay shifts the vector $2S$-wave state approximately by 65~MeV and insignificantly broadens  the peak:

\begin{equation}
\Delta \tilde{M}_{2S}  \approx 2 \frac{\Delta M \sqrt{\Delta M''^2 - M_{2\pi}^2}}{M}\approx 10~\textrm{MeV},
\end{equation}
where
$\Delta M= M(B_c^*)-M(B_c)$ and
$\Delta M''= M(B_c(2 ^3S_1))-M(B_c^*)$. 
As a result, the visible peak from the vector $2 ^3S_1$ state  in $B_c+\pi\pi$ spectrum will appear  30~MeV 
to the left  from  the pseudoscalar $2 ^1S_0$ state, in spite of that $2 ^3S_1$ excitation is about 40~MeV heavier, than  $2 ^1S_0$ one (see the solid histogram in fig.~\ref{fig:2s}).

The apparatus resolution is taken into account by convolution of this distribution with a Gaussian function with a dispersion $15$~GeV. It could be concluded, that if two these peaks will be resolved in the experiment due to small distance, a single asymmetric peak will be seen (the  dashed histogram in fig.~\ref{fig:2s}).

\section{Conclusion}

It is shown in this work that the loss of ``soft'' photon from the decay $B_c^*\to B_c +\gamma$  shifts the peaks of $2S$-, $2P$- and $3P$-excitations 
by $~65$~MeV and broadens them by $10-20$~MeV. This doesn't preclude the study of $B_c$ excitations. 

In  early theoretical works on the hadronic $B_c$-meson production it was shown that the  ratio $R_{B_c} = \sigma(B_c^*)/\sigma(B_c)$ could provide an essential information about the $B_c$ production mechanism. 
Unfortunately, is it extremely difficult to measure this ratio  due to a small energy release in the $B_c^* \to B_c \gamma$ decay. 
Most likely,  $2P$-wave excitations  will be found first in the $M_{B_c}+\gamma$ spectrum.

It is also shown, that decays $B_c(2S) \to B_c(B_c^*)+\pi^+\pi^-$ decays may appear to be a  source of new information about the nature of $\sigma$-meson.

\bibliographystyle{ieeetr}
\bibliography{bc_pwave}

\begin{thebibliography}{10}

\bibitem{Aaltonen:2007gv}
T.~Aaltonen {\em et~al.}, ``{Observation of the Decay $B^+$ -($c$) $\to J/\psi
  \pi^\pm$ and Measurement of the $B^+$ -($c$) Mass},'' {\em Phys.Rev.Lett.},
  vol.~100, p.~182002, 2008.

\bibitem{Abazov:2008rba}
V.~Abazov {\em et~al.}, ``{Measurement of the lifetime of the $B_c^\pm$ meson
  in the semileptonic decay channel},'' {\em Phys.Rev.Lett.}, vol.~102,
  p.~092001, 2009.

\bibitem{Abazov:2008kv}
V.~Abazov {\em et~al.}, ``{Observation of the $B_c$ Meson in the Exclusive
  Decay $B_c \to J/\psi \pi$},'' {\em Phys.Rev.Lett.}, vol.~101, p.~012001,
  2008.

\bibitem{Abe:1998wi}
F.~Abe {\em et~al.}, ``{Observation of the $B_c$ meson in $p\bar{p}$ collisions
  at $\sqrt{s} = 1.8$ TeV},'' {\em Phys.Rev.Lett.}, vol.~81, pp.~2432--2437,
  1998.

\bibitem{Aaij:2012dd}
R.~Aaij {\em et~al.}, ``{Measurements of $B_c^+$ production and mass with the
  $B_c^+ \to J/\psi \pi^+$ decay},'' {\em Phys.Rev.Lett.}, vol.~109, p.~232001,
  2012.

\bibitem{CMS:2012oxa}
W.~Adam {\em et~al.}, ``{Observation of the decays Bc to J/psi pi and Bc to
  J/psi pi pi pi in pp collisions at sqtr(s) = 7 TeV},'' {\em
  CMS-PAS-BPH-11-003}, 2012.

\bibitem{ATLAS:2012bja}
G.~Aad {\em et~al.}, ``{Observation of the $B^{\pm}_c$ meson in the decay
  $B^{\pm}_c\to J/\psi(\mu^+\mu^-)\pi^{\pm}$ with the ATLAS detector at the
  LHC},'' {\em ATLAS-CONF-2012-028, ATLAS-COM-CONF-2012-035}, 2012.

\bibitem{LHCb:2012ag}
R.~Aaij {\em et~al.}, ``{First observation of the decay $B_c^+ \to J/\psi
  \pi^+\pi^-\pi^+$},'' {\em Phys.Rev.Lett.}, vol.~108, p.~251802, 2012.

\bibitem{Aaij:2013oya}
R.~Aaij {\em et~al.}, ``{Observation of the decay $B_c^+ \to \psi(2S)\pi^+$},''
  {\em Phys.Rev.}, vol.~D87, p.~071103, 2013.

\bibitem{Aaij:2013gia}
R.~Aaij {\em et~al.}, ``{Observation of $B^+_c \rightarrow J/\psi D_s^+$ and
  $B^+_c \rightarrow J/\psi D_s^{*+}$ decays},'' {\em Phys.Rev.}, vol.~D87,
  p.~112012, 2013.

\bibitem{Aaij:2013vcx}
R.~Aaij {\em et~al.}, ``{First observation of the decay $B_{c}^{+}\to J/\psi
  K^+$},'' 2013.

\bibitem{Gershtein:1994jw}
S.~Gershtein, V.~Kiselev, A.~Likhoded, and A.~a. Tkabladze, ``{Physics of B(c)
  mesons},'' {\em Phys.Usp.}, vol.~38, pp.~1--37, 1995.

\bibitem{Gershtein:1997qy}
S.~Gershtein, V.~Kiselev, A.~Likhoded, A.~Tkabladze, A.~Berezhnoy, {\em
  et~al.}, ``{Theoretical status of the B(c) meson},'' 1997.

\bibitem{Gouz:2002kk}
I.~Gouz, V.~Kiselev, A.~Likhoded, V.~Romanovsky, and O.~Yushchenko,
  ``{Prospects for the $B_c$ studies at LHCb},'' {\em Phys.Atom.Nucl.},
  vol.~67, pp.~1559--1570, 2004.

\bibitem{Godfrey:2004ya}
S.~Godfrey, ``{Spectroscopy of $B_c$ mesons in the relativized quark model},''
  {\em Phys.Rev.}, vol.~D70, p.~054017, 2004.

\bibitem{Berezhnoy:1994ba}
A.~Berezhnoy, A.~Likhoded, and M.~Shevlyagin, ``{Hadronic production of B(c)
  mesons},'' {\em Phys.Atom.Nucl.}, vol.~58, pp.~672--689, 1995.

\bibitem{Chang:1994aw}
C.-H. Chang, Y.-Q. Chen, G.-P. Han, and H.-T. Jiang, ``{On hadronic production
  of the B(c) meson},'' {\em Phys.Lett.}, vol.~B364, pp.~78--86, 1995.

\bibitem{Berezhnoy:1995au}
A.~Berezhnoy, A.~Likhoded, and O.~Yushchenko, ``{Some features of the hadronic
  B*(c) meson production at large p(T)},'' {\em Phys.Atom.Nucl.}, vol.~59,
  pp.~709--713, 1996.

\bibitem{Kolodziej:1995nv}
K.~Kolodziej, A.~Leike, and R.~Ruckl, ``{Production of B(c) mesons in hadronic
  collisions},'' {\em Phys.Lett.}, vol.~B355, pp.~337--344, 1995.

\bibitem{Berezhnoy:1996ks}
A.~Berezhnoy, V.~Kiselev, and A.~Likhoded, ``{Hadronic production of S and P
  wave states of anti-b c quarkonium},'' {\em Z.Phys.}, vol.~A356, pp.~79--87,
  1996.

\bibitem{Berezhnoy:1997fp}
A.~Berezhnoy, V.~Kiselev, A.~Likhoded, and A.~Onishchenko, ``{B(c) meson at
  LHC},'' {\em Phys.Atom.Nucl.}, vol.~60, pp.~1729--1740, 1997.

\bibitem{Baranov:1997sg}
S.~Baranov, ``{Semiperturbative and nonperturbative production of hadrons with
  two heavy flavors},'' {\em Phys.Rev.}, vol.~D56, pp.~3046--3056, 1997.

\bibitem{Baranov:1997wv}
S.~Baranov, ``{Single and pair production of B/c mesons in p p, e p, and gamma
  gamma collisions},'' {\em Phys.Atom.Nucl.}, vol.~60, pp.~1322--1332, 1997.

\bibitem{Berezhnoy:1997uz}
A.~Berezhnoy, V.~Kiselev, and A.~Likhoded, ``{Hadroproduction of the S- and P
  wave states of anti-b c quarkonium},'' {\em Phys.Atom.Nucl.}, vol.~60,
  pp.~100--109, 1997.

\bibitem{Chang:2004bh}
C.-H. Chang, J.-X. Wang, and X.-G. Wu, ``{Hadronic production of the P-wave
  excited $B_c$ -states B*(cJ, L=1)},'' {\em Phys.Rev.}, vol.~D70, p.~114019,
  2004.

\bibitem{Berezhnoy:2004gc}
A.~Berezhnoy, ``{Color flows for the process $gg\to B_c + c + \bar b$},'' {\em
  Phys.Atom.Nucl.}, vol.~68, pp.~1866--1872, 2005.

\bibitem{Chang:2005wd}
C.-H. Chang, C.-F. Qiao, J.-X. Wang, and X.-G. Wu, ``{Hadronic production of
  $B_c (B^*_c)$ meson induced by the heavy quarks inside the collision
  hadrons},'' {\em Phys.Rev.}, vol.~D72, p.~114009, 2005.

\bibitem{Chang:2006xka}
C.-H. Chang, J.-X. Wang, and X.-G. Wu, ``{An Upgraded version of the generator
  BCVEGPY2.0 for hadronic production of B(c)meson and its excited states},''
  {\em Comput.Phys.Commun.}, vol.~175, pp.~624--627, 2006.

\bibitem{Berezhnoy:2010wa}
A.~Berezhnoy, A.~Likhoded, and A.~Martynov, ``{Associative production of $B_c$
  and $D$ mesons at LHC},'' {\em Phys.Rev.}, vol.~D83, p.~094012, 2011.

\bibitem{Gupta:1995ps}
S.~N. Gupta and J.~M. Johnson, ``{B(c) spectroscopy in a quantum chromodynamic
  potential model},'' {\em Phys.Rev.}, vol.~D53, pp.~312--314, 1996.

\bibitem{Kiselev:1994rc}
V.~Kiselev, A.~Likhoded, and A.~Tkabladze, ``{B(c) spectroscopy},'' {\em
  Phys.Rev.}, vol.~D51, pp.~3613--3627, 1995.

\bibitem{Brown:1975dz}
L.~S. Brown and R.~N. Cahn, ``{Chiral Symmetry and psi-prime --- psi + pi + pi
  Decay},'' {\em Phys.Rev.Lett.}, vol.~35, p.~1, 1975.

\bibitem{Novikov:1980fa}
V.~Novikov and M.~A. Shifman, ``{Comment on the psi-prime --- J/psi pi pi
  Decay},'' {\em Z.Phys.}, vol.~C8, p.~43, 1981.

\bibitem{Voloshin:1975yb}
M.~B. Voloshin, ``{Adler's Selfconsistency Condition in the Decay psi-prime
  (3700) --- psi (3100) pi pi},'' {\em JETP Lett.}, vol.~21, pp.~347--348,
  1975.

\bibitem{Voloshin:1980zf}
M.~B. Voloshin and V.~I. Zakharov, ``{Measuring QCD Anomalies in Hadronic
  Transitions Between Onium States},'' {\em Phys.Rev.Lett.}, vol.~45, p.~688,
  1980.

\bibitem{Ablikim:2006bz}
M.~Ablikim {\em et~al.}, ``{Production of sigma in psi(2S) -> pi+ pi- J/psi},''
  {\em Phys.Lett.}, vol.~B645, pp.~19--25, 2007.

\bibitem{Albaladejo:2012te}
M.~Albaladejo and J.~Oller, ``{On the size of the sigma meson and its
  nature},'' {\em Phys.Rev.}, vol.~D86, p.~034003, 2012.

\bibitem{Black:2000qq}
D.~Black, A.~H. Fariborz, S.~Moussa, S.~Nasri, and J.~Schechter, ``{Unitarized
  pseudoscalar meson scattering amplitudes in three flavor linear sigma
  models},'' {\em Phys.Rev.}, vol.~D64, p.~014031, 2001.

\bibitem{Harada:2012km}
M.~Harada, H.~Hoshino, and Y.-L. Ma, ``{Effect of sigma meson on the $D_1(2430)
  \to D\pi\pi$ decay},'' {\em Phys.Rev.}, vol.~D85, p.~114027, 2012.

\end{thebibliography}

\end{document}